\documentclass[aps,amsmath,showpacs,showkeys,superscriptaddress,10pt]{revtex4}
\usepackage{color}
\usepackage{amssymb}
\usepackage{titlesec}
\usepackage{amssymb,amsfonts,amsthm}
\begin{document}
\title{On the analytic representation of Newtonian systems}
\author{BENOY TALUKDAR}
\email{binoyt123@rediffmail.com}
\affiliation{Department of Physics, Visva-Bharati University, Santiniketan 731235, India}
\author{SUPRIYA CHATTERJEE}
\affiliation{Department of Physics, Bidhannagar College, EB-2, Sector-1, Salt Lake, Kolkata 700064, India}
\author{SEKH GOLAM ALI} 
\affiliation{Department of Physics, Kazi Nazrul University, Asansol 713303, India}
\begin{abstract}
We show that the theory of self-adjoint differential equations can be used to provide a satisfactory solution of the inverse variational problem in classical mechanics.  A Newtonian equation when transformed to the self-adjoint form allows one to find an appropriate Lagrangian representation (direct analytic representation) for it. On the other hand, the same Newtonian equation in conjunction with its adjoint provides a basis to construct a different Lagrangian representation (indirect analytic representation) for the system. We obtain the time-dependent Lagrangian of the damped Harmonic oscillator from the self-adjoint form of the equation of motion and at the same time identify the adjoint of the equation with the so called Bateman image equation with a view to construct a time-independent indirect Lagrangian representation. We provide a number of case studies to demonstrate the usefulness of the approach derived by us. We also present similar results for a number of nonlinear differential equations by using an integral representation of the Lagrangian function and make some useful comments.
\end{abstract}
\maketitle
\section*{1. Introduction}
In point mechanics the term \textquoteleft analytic representation' refers to description of Newtonian systems by means of Lagrangians \cite{1}. Understandably, to find the analytic representation of a mechanical system one begins with the equation of motion and then constructs a Lagrangian function by using a strict mathematical procedure discovered by Helmholtz \cite{2, 3}. In the calculus of variation this is the so-called inverse variational problem which is more complicated than the usual direct problem where one first assigns a Lagrangian function using phenomenological consideration and then computes the equation of motion using the Euler-Lagrange equation \cite{4}. However, there are two types of analytic representation, namely, the direct and indirect ones. We can introduce the basic concepts of direct and indirect analytic representations by using a system of two uncoupled harmonic oscillators with equations of motion 
\begin{equation}
\ddot{q}(t)+\omega^2 q(t)=0  
\end{equation}
and 
\begin{equation}
\ddot{y}(t)+\omega^2 y(t)=0. 
\end{equation}
It is straightforward to verify that the system of equations (1) and (2) can be analytically represented either by the Lagrangian
\begin{equation}
 L_d=\frac{1}{2}(\dot{q}^2(t)+\dot{y}^2(t))-\frac{\omega^2}{2}(q^2(t)+y^2(t))
\end{equation}
or by the Lagrangian
\begin{equation}
 L_i=\dot{q}(t)\dot{y}(t)-\omega^2q(t)y(t).
\end{equation}
Here overdots denote differentiation with respect to time $t$. The function $L_d$ refers to a Lagangian that gives the direct analytic representation of the system presumably because it yields the equation of motion for $q(t)\;(y(t))$ via the Euler-Lagrange equation written in terms of $q(t)\;(y(t))$. On the other hand, $L_i$ yields the equation of motion for  $q(t)\;(y(t))$ via the Euler-Lagrange equation written in terms of $y(t)\;(q(t))$. This is why the representation of the system by the use of $L_i$ is called indirect analytic representation. This simple example indicates that Lagrangian representations of Newtonian systems are not unique. The problem of non-uniqueness of the Lagrangian functions has deep consequences for the correspondence between symmetries and constants of the motion. For example, the direct Lagrangian (3) is rotationally invariant such that the associated Noether constant of the motion is the angular momentum. As opposed to this, the indirect Lagrangian (4) is invariant under 'squeeze' transformation $(q(t),\;y(t)\rightarrow q(t)e^t,\;y(t)e^{-t})$. Consequently, for the Lagrangian $L_i$, conservation of angular momentum is associated with the invariance under squeeze \cite{5}.
\par In this work we shall first examine how Hemholtz conditions are useful to study analytic representation of Newtonian systems modeled by linear second-order ordinary differential equations and then provide a general method to construct their direct and indirect analytic representations by using the theory of self-adjoint differential equations \cite{6}. In particular, we show that the self-adjoint form of a Newtonian equation can always be used to find its direct analytic representation even if the original equation does not satisfy the Helmholtz criteria \cite{2,3}. On the other hand, any Newtonian equation in conjunction with its adjoint permits one to construct its indirect analytical representation. We shall demonstrate the simplicity and effectiveness of our approach by presenting a number of case studies.  
\par The general concept of self-adjointness for linear differential equations is well documented in the mathematical literature \cite{7, 8}. This is, however, not the case with the nonlinear  equations although there have been some attempts to build a theory of nonlinear self-adjointness \cite{9, 10}. It appears that it will not be straightforward to use this theory to study the analytic representation of nonlinear systems. However, in the recent past the nonlinear variational problem has been studied \cite{11,12,13} employing an integral representation \cite{14} of the Lagrangian function. The integral representation was derived by applying the Cauchy's method of characteristics \cite{15} to solve the equation satisfied by the first integral of an N-dimensional autonomous system. From (1), (2) and (4) it is obvious that the indirect analytic representation of a linear system is obtained by doubling the number of degrees of freedom. It will be quite significant to examine if the same is also true for uncoupled nonlinear equations. Admittedly, a straightforward way to achieve this consists in using the integral representation of the Lagrangian function sought in ref. 14.
\par In Sec. 2 we indicate how one can find the Lagrangian representation of a Newtonian system when its equation of motion satisfies Helmholtz condition. We adapt, in Sec.3, the theory of self-adjoint second-order linear differential equations to provide a complete solution for the inverse problem of the calculus of variation. With special attention to the damped Harmonic oscillator we obtain both direct and indirect analytic representation of the system. We devote Sec. 4 to present results for direct and indirect analytic representation for some linear second-order ordinary differential equations of mathematical physics. In Sec. 5 we consider a two dimensional autonomous system (linear or nonlinear), provide an integral representation for its Lagrangian function and subsequently, make use of the result to derive analytic representations of a number of nonlinear Newtonian equations. We find some differences between the analytic representations of uncoupled linear and nonlinear equations. The coupled nonlinear systems, however, exhibit properties which are formally similar to those of the corresponding linear equations. Finally, in Sec. 6 we summarize our outlook on the present work and make some concluding remarks.
\section*{2. Helmholtz conditions and Lagrangian representation}
From the inverse problems in the calculus of variations \cite{1} one knows that all Newtonian systems cannot have Lagrangian representation. In particular, the equations written in the general form
\begin{equation}
 F_i=A_{ij}(t,q,\dot{q})\ddot{q}^j+B_i(t,q,\dot{q})=0,\;\;\;\;\;i,j=1,\,2,.....,\,n \;\;\;\mbox{and}\;\;\; q=q(t)\in R^n
\end{equation}
will have a Lagrangian representation if and only if
\begin{subequations}
\begin{equation}
\frac{\partial F_i}{\partial \ddot{q}^j}=\frac{\partial F_j}{\partial \ddot{q}^i},
\end{equation}
\begin{equation}
\frac{\partial F_i}{\partial \dot{q}^j}+\frac{\partial F_j}{\partial \dot{q}^i}=2\frac{d}{dt}(\frac{\partial F_i}{\partial \ddot{q}^j}) 
\end{equation}
\mbox{and}
\begin{equation}
 \frac{\partial F_i}{\partial q^j}-\frac{\partial F_j}{\partial q^i}=\frac{1}{2}\frac{d}{dt}(\frac{\partial F_i}{\partial \dot{q}^j}-\frac{\partial F_j}{\partial \dot{q}^i}).
\end{equation}
\end{subequations}
The relations (6a)-(6c) are often called Helmholtz conditions \cite{2,3} and give the necessary and sufficient conditions for the existence of a Lagrangian function for any Newtonian system. Equation (5) represents an $n$ dimensional differential equation. For the one-dimensional case (6a) and (6c) become identity such that we are now left with only one condition
\begin{equation}
 \frac{\partial F}{\partial\dot{q}}=\frac{d}{dt}(\frac{\partial F}{\partial \ddot{q}}).
\end{equation}
The general equation for the one-dimensional Newtonian system (linear) can be written as
\begin{equation}
 F=\ddot{q}(t)+r(t)\dot{q}(t)+s(t)q(t)=0.
\end{equation}
Equation (8) will satisfy the Helmholtz condition (7) if $r(t)=0$. Thus we have
\begin{equation}
 \ddot{q}+s(t)q(t)=0.
\end{equation}
Multiplying (9) by $\delta q$ and integrating over $t$ from $t_1$ to $t_2$ we can recast it in the variational form
\begin{equation}
 \delta\int_{t_1}^{t_2}(\frac{1}{2}\dot{q}^2(t)-\frac{s(t)}{2}q^2(t))dt=0
\end{equation}
such that (9) follows from the Lagrangian function
\begin{equation}
 L=\frac{1}{2}\dot{q}^2(t)-\frac{s(t)}{2}q^2(t)
\end{equation}
via the Euler-Lagrange equation
\begin{equation}
 \frac{d}{dt}(\frac{\partial L}{\partial \dot{q}(t)})-\frac{\partial L}{\partial q(t)}=0.
\end{equation}
\section*{3. Inverse problem using the theory of self-adjoint differential equations}
Here we shall take recourse to the use of the theory for self-adjoint differential equations to construct Lagrangian representation of Newtonian systems. For arbitrary values of $r(t)$ and $s(t)$, (8) is not self-adjoint. However, the theory of linear second-order self-adjoint differential equations is quite general \cite{5}. For example, we can always transform (8) in the self-adjoint form $F_{sadj}$ by multiplying it with a non-vanishing factor 
\begin{equation}
 \rho(x)=e^{\int r(t)dt}
\end{equation}
such that
\begin{equation}
 F_{sadj}=\frac{d}{dt}(\dot{q}(t)e^{\int r(t)dt})+s(t)q(t)e^{\int r(t)dt}=0.
\end{equation}
On the other hand, the adjoint equation  $F_{adj}$  corresponding to (8) can be found by changing the dependent variable by
\begin{equation}
 q(t)=y(t)e^{-\int r(t)dt}.
\end{equation}
We thus have
\begin{equation}
 F_{adj}=\ddot{y}(t)+b(t)\dot{y}(t)+c(t)y(t)=0
\end{equation}
with
\begin{equation}
 b(t)=-r(t)\;\;\;\mbox{and} \;\;\;c(t)=-\dot{r}(t)+s(t).
\end{equation}
Understandably, the equation will be self-adjoint if $F=F_{adj}$. 
\par Multiplying (14) by $\delta q(t)$ and integrating over $t$ we can recast it in the variational form
\begin{equation}
 \delta\int_{t_1}^{t_2}L_d(q(t),\;\dot{q}(t),\;t)dt=0
\end{equation}
such that
\begin{equation}
 L_d=e^{\int r(t)dt}(\frac{\dot{q}^2(t)}{2}-s(t)\frac{q^2(t)}{2}).
\end{equation}
We now make use of (8) and its adjoint (16) to construct a Lagrangian for the system as
\begin{equation}
 L_i=y(t)F+q(t)F_{adj}-\frac{d}{dt}(y(t)\dot{q}(t))-\frac{d}{dt}(q(t)\dot{y}(t)).
\end{equation}
In writing (20) we have assumed that the Lagrangian of a system is its own equation of motion \cite{4,11}. The third and fourth terms represent the trivial gauge terms \cite{16,17} for a second-order Lagrangian and have been introduced only to write a first-order Lagrangian for the system. Equation (20) on simplification reads
\begin{equation}
 L_i=\dot{q}(t)\dot{y}(t)+\frac{r(t)}{2}(q(t)\dot{y}(t)-y(t)\dot{q}(t))+(\frac{1}{2}\dot{r}(t)-s(t))q(t)y(t).
\end{equation}
It is now straightforward to verify that the explicitly time-dependent Lagrangian $L_d$ in (19) gives a direct analytic representation of (8), while the explicitly time-independent Lagrangian $L_i$ in (21) gives its indirect analytic representation. This confirms that the theory of self-adjoint differential equations provides us with a unique mathematical framework to study the Lagrangian structure of any Newtonian system modeled by general linear second-order ordinary differential equation which may or may not satisfy the Helmholtz condition \cite{2,3}. As a very instructive example for the effectiveness of the approach derived, we first consider the variational properties for the linearly damped Harmonic oscillator.
\par The equation of motion for the damped Harmonic oscillator can be written as 
\begin{equation}
 F=m\ddot{q}(t)+\gamma\dot{q}(t)+kq(t)=0,
\end{equation}
where $m$ and $k$ stand for the mass and spring constant of the oscillator and the symbol $\gamma$ represents the frictional coefficient of the medium in which the oscillation takes place. The term $\gamma\dot{q}(t)$ accounts for the dissipation of energy from the system to the environment. The dissipative or non-conservative systems were found not to follow naturally from the Hamilton's variational principle. Consequently, there have been long standing efforts to construct action functionals for such systems. The first proposition in respect of this was made by Rayleigh \cite{4} who introduced a dissipation function $D=\gamma\dot{x}^2(t)$ in addition to the usual Harmonic oscillator Lagrangian $L$ to express (22) in the variational form. This proposition did not receive much attention presumably because, rather than one, here we require two scalar functions, namely $L$ and $D$, to write the action principle. Other major propositions include (a) construction of a time-dependent Lagrangian \cite{18,19,20} directly from (22) and (b) construction of an explicitly time-independent Lagrangian by doubling the degrees of freedom of the system \cite{21}. Here we shall obtain these time-dependent and time-independent analytic representations by using the theory of self-adjoint differential equations alone.
\par From (8) and (22) $r(t)=\gamma/m$ and $s(t)=k/m$. Using these values in (14) and (19) we write 
\begin{equation}
 F_{sadj}=e^{\frac{\gamma t}{m}}(\ddot{q}(t)+\frac{\gamma}{m}\dot{q}(t)+\frac{k}{m}q(t))=0
\end{equation}
and the Lagrangian 
\begin{equation}
 L_d=\frac{e^{{\frac{\gamma t}{m}}}}{m}(\frac{1}{2}m\dot{q}^2(t)-\frac{1}{2}kq^2(t))
\end{equation}
for the direct analytic representation of the damped Harmonic oscillator. Equation (24) represents the result first reported  in ref. 18. 
We shall now use (22) and the corresponding adjoint equation to obtain a time-independent Lagrangian for the damped Harmonic oscillator. Meanwhile, it will be useful to provide a brief review of the work in ref. 21 in which Bateman considered the damped oscillator together with an amplified oscillator such that the energy drained out from the first is completely absorbed by the second.  Understandably, we have now a dual system which is closed. The amplified oscillator associated with the damped system (22) was chosen as
\begin{equation}
 F'=m\ddot{y}(t)-\gamma\dot{y}(t)+ky(t)=0.
\end{equation}
Mechanistically, this attempt to understand dissipation by the simultaneous use of (22) and (25) amounts to doubling the degrees of freedom to study the problem. Using $F$ and $F'$ we write a Lagrangian given by
\begin{equation}
 L_i=(y(t)F+q(t)F'-m\frac{d}{dt}(y(t)\dot{q}(t)+q(t)\dot{y}(t)))/2.
\end{equation}
In the simplified form (26) reads
\begin{equation}
 L_i=m\dot{q}(t)\dot{y}(t)+\frac{\gamma}{2}(q(t)\dot{y}(t)-\dot{q}(t)y(t))-kq(t)y(t).
\end{equation}
The explicitly time-independent result in (27) provides an indirect analytic representation of the damped harmonic oscillator. 
\par From the above it is evident that in order to bring the damped Harmonic oscillator within the framework of variational principle, the image equation (25) was introduced by Bateman \cite{21} using purely phenomenological arguments. It is an interesting mathematical curiosity to note that (25) represents the adjoint equation of (22). This can easily be proved by making use of (16) and (17). Thus we see that, from mathematical point of view, the amplified oscillator of Bateman is represented by the adjoint of the damped Harmonic oscillsator.
\par From the self-adjoint forms of (22) and (25) it is also possible to find a direct Lagrangian
\begin{equation}
L_d=\frac{e^{{\frac{\gamma t}{m}}}}{m}(\frac{1}{2}m\dot{q}^2(t)-\frac{1}{2}kq^2(t))+\frac{e^{{-\frac{\gamma t}{m}}}}{m}(\frac{1}{2}m\dot{y}^2(t)-\frac{1}{2}ky^2(t)) 
\end{equation}
for the Bateman dual system.
\par We conclude by noting that (8) with prescribed analytical properties of the real valued coefficients $r(t)$ and $s(t)$ over some region of interest $a\leq t\leq b$ represents the most general linear second-order homogeneous ordinary differential equation. The self-adjoint form of the equation leads rather naturally to its direct analytic representation. On the other hand, the indirect analytic representation of the equation can be obtained by combining it with the associated adjoint equation. 
\section*{4. Analytic representation of some special second-order differential equations}
Here we make use of formalism of Sec. 3 to obtain results for direct and indirect Lagrangians of a number of equations which have important applications in physical theories. In Tables I and II we present results for seven such equations. In particular, Table I gives the results for direct Lagrangian while Table II contains similar results for the indirect Lagrangian. In presenting the results we always use $t$ as the independent variable and, for brevity, call it as time. Column 2 in Table I gives the self-adjoint equations corresponding to the original equations in column 1.The expressions for the Lagrangian are presented in column 3.
\begin{table}[h]
\begin{center}
\begin{tabular}{|l|l|l|}
\hline
Original differential equation & Self-adjoint equation & Lagrangian giving direct \\
 & & analytic representation\\
\hline
Legendre:& & \\
$(1-t^2)\ddot{q}(t)-2t\dot{q}(t)+n(n+1)q(t)=0$ &$(1-t^2)\ddot{q}(t)-2t\dot{q}(t)+n(n+1)q(t)=0$ &$\frac{1}{2}(1-t^2)\dot{q}^2(t)-\frac{1}{2}n(n+1)q^2(t)$ \\
\hline
Bessel:& & \\
$t^2\ddot{q}(t)+t\dot{q}(t)+(t^2-\nu^2)q(t)=0$ &$t\ddot{q}(t)+\dot{q}(t)+(t-\frac{\nu^2}{t})q(t)=0$&$\frac{1}{2}t\dot{q}^2(t)-\frac{1}{2}(t-\frac{\nu^2}{t})q^2(t)$  \\
\hline
Laguerre:& & \\
$t\ddot{q}(t)+(1-t)\dot{q}(t)+nq(t)=0$&$e^{-t}(t\ddot{q}(t)+(1-t)\dot{q}(t)+nq(t))=0$& $\frac{1}{2}te^{-t}\dot{q}^2(t)-\frac{1}{2}ne^{-t}q^2(t)$ \\
\hline
Hermite: & & \\
$\ddot{q}(t)-2t\dot{q}(t)+2nq(t)=0$ & $e^{-t^2}(\ddot{q}(t)-2t\dot{q}(t)+2nq(t))=0$ & $\frac{1}{2}e^{-t^2}\dot{q}^2(t)-ne^{-t^2}q^2(t)$\\
\hline
Chebyshev:& & \\
$(1-t^2)\ddot{q}(t)-t\dot{q}(t)+n^2q(t)=0$ & $\frac{-1}{\sqrt{t^2-1}}((1-t^2)\ddot{q}(t)-t\dot{q}(t)+n^2q(t))=0$ & $\frac{1}{2}\sqrt{t^2-1}\dot{q}^2(t)+\frac{1}{2\sqrt{t^2-1}}n^2q^2(t)$\\
\hline
Gaussian Hypergeometric: & & \\
$(1-t)t\ddot{q}(t)+((1+\alpha+\beta)t-\gamma)\dot{q}(t)$&$(1-t)^{-2-\alpha-\beta+\gamma}t^{-1-\gamma}((1-t)t\ddot{q}(t)+$ & $t^{-\gamma}(1-t)^{-1+\alpha+\beta+\gamma}(\frac{1}{2}\dot{q}^2(t)$\\
$-\alpha\beta q(t)=0$ &$((1+\alpha+\beta)t-\gamma)\dot{q}(t)-\alpha\beta q(t))=0$ &$+\frac{\alpha\beta q^2(t)}{2t(1-t)})$ \\
\hline
Confluent Hypergeometric: & & \\
$t\ddot{q}(t)+(\gamma-t)\dot{q}(t)-\alpha q(t)=0$& $e^{-t}t^{c-1}(t\ddot{q}(t)-(\gamma-t)\dot{q}(t)-\alpha q(t))=0$& $t^{\gamma}e^{-t}(\frac{1}{2}\dot{q}^2(t)+\frac{1}{2t}\alpha q^2(t))$\\
\hline
\end{tabular}
\caption{Self-adjoint differential equations and  Lagrangians giving direct analytic representation of some important linear second-order differential equation of mathematical physics.}
\end{center}
\end{table}
The Legendre equation is self-adjoint such that the equations in columns 1 and 2 of row 1 are same. On the other hand, the other equations in the table are non-self-adjoint. Consequently, for these equations the self-adjoint forms are different from the parent equations. All Lagrangians giving the direct analytic representation are explicitly time dependent and closely resemble the result in (24) for the damped harmonic oscillator. 
\par In close analogy with the results displayed in Table I we reserve columns 1 and 2 of Table II for the original equations and their adjoints. In column 3 we present results for Lagrangians giving indirect analytic representation for the same set of equations as considered in Table I. Looking closely into the entries of Table II we see that the Legendre equation and its adjoint are same. This result is quite expected since Legendre equation is self-adjoint. The Lagrangian function for the Legendre equation is of the same form as that in (27) for the damped harmonic oscillator except that the Lagrangian does not involve any term analogous to the middle term in (27). This is, however, not true for other equations in the table, which are not self-adjoint. For example, the Lagrangian functions for all other equations have middle terms in the form $\beta(t)=y(t)\dot{q}(t)-q(t)\dot{y}(t)$.
\begin{table}[h]
\begin{center}
\begin{tabular}{|l|l|l|}
\hline
Original differential equation & Adjoint equation & Lagrangian giving indirect \\
 & & analytic representation\\
\hline
Legendre:& & \\
$(1-t^2)\ddot{q}(t)-2t\dot{q}(t)+n(n+1)q(t)=0$ &$(1-t^2)\ddot{y}(t)-2t\dot{y}(t)+n(n+1)y(t)=0$ &$(1-t^2)\dot{q}(t)\dot{y}(t)-n(n+1)q(t)y(t)$ \\
\hline
Bessel:& & \\
$t^2\ddot{q}(t)+t\dot{q}(t)+(t^2-\nu^2)q(t)=0$ &$t^2\ddot{y}(t)+3t\dot{y}(t)+(1+t^2-\nu^2)y(t)=0$&$t^2\dot{q}(t)\dot{y}(t)+\frac{1}{2}t(y(t)\dot{q}(t)-q(t)\dot{y}(t))$  \\
 & & $-(t^2-\nu^2+\frac{1}{2})q(t)y(t)$ \\
\hline
Laguerre:& & \\
$t\ddot{q}(t)+(1-t)\dot{q}(t)+nq(t)=0$&$t\ddot{y}(t)+(1+t)\dot{y}(t)+n(n+1)y(t)=0$& $t\dot{q}(t)\dot{y}(t)+\frac{1}{2}t(y(t)\dot{q}(t)-q(t)\dot{y}(t))$ \\
 & & $-(n+\frac{1}{2})q(t)y(t)$\\
\hline
Hermite: & & \\
$\ddot{q}(t)-2t\dot{q}(t)+2nq(t)=0$ & $\ddot{y}(t)+2t\dot{y}(t)+2(n+1)y(t)=0$ & $\dot{q}(t)\dot{y}(t)+t(y(t)\dot{q}(t)-q(t)\dot{y}(t))$\\
 & & $-(2n+1)q(t)y(t)$\\
\hline
Chebyshev:& & \\
$(1-t^2)\ddot{q}(t)-t\dot{q}(t)+n^2q(t)=0$ & $(1-t^2)\ddot{y}(t)-3t\dot{y}(t)-$ & $(1-t^2)\dot{q}(t)\dot{y}(t)+$\\
 &$(1-n^2)y(t)=0$ & $\frac{1}{2}t(q(t)\dot{y}(t)-y(t)\dot{q}(t))$\\
 & & $+(\frac{1}{2}-n^2)q(t)y(t)$\\
\hline
Gaussian Hypergeometric: & & \\
$(1-t)t\ddot{q}(t)+((1+\alpha+\beta)t-\gamma)\dot{q}(t)$&$(1-t)t\ddot{y}(t)+(2-t(5+\alpha+\beta)+\gamma)\dot{y}(t)$ & $(1-t)t\dot{q}(t)\dot{y}(t)+$\\
$-\alpha\beta q(t)=0$ &$-(3+\alpha+\beta+\alpha\beta)y(t)=0$ &$\frac{1}{2}(1-3t-\alpha t-\beta t+\gamma)\times$ \\
 & & $(\dot{q}(t)y(t)-\dot{y}(t)q(t))+$\\
 & & $\frac{1}{2}(3+\alpha+\beta+\alpha\beta)q(t)y(t)$\\
\hline
Confluent Hypergeometric: & & \\
$t\ddot{q}(t)+(\gamma-t)\dot{q}(t)-\alpha q(t)=0$& $t\ddot{y}(t)+(2+t-\gamma)\dot{y}(t)+$& $t\dot{q}(t)\dot{y}(t)+\frac{1}{2}(1+t-\gamma)\times$\\
 &$(1-\alpha)y(t)=0$ & $(y(t)\dot{q}(t)-q(t)\dot{y}(t))-$\\
 & & $(\frac{1}{2}-\alpha)q(t)y(t)$\\
\hline
\end{tabular}
\caption{Adjoint differential equations and Lagrangians giving indirect analytic representation of the linear second-order differential equations in Table I.}
\end{center}
\end{table}
\par All examples in Tables I and II represent homogeneous differential equations. It will be instructive to examine if the corresponding non-homogeneous equations are also Lagrangian. This is important because the non-homogeneous term incorporates the effects of the external force on the physical system. Moreover, the equations  in the tables also play a crucial role in off-energy-shell potential scattering \cite{22,23,24}. In the following we make use of a simple system to demonstrate that the non-homogeneity of differential equations does not bring in any serious complications to construct their analytic representation.
\par We are interested in a forced oscillator whose  equation of motion is given by
\begin{equation}
 \ddot{q}(t)+\omega^2q(t)=F(t),
\end{equation}
where $F(t)$ stands for an external force acting on the system. It is straightforward to transform (29) in the variational form so as to obtain the Lagrangian
\begin{equation}
 L_d=\frac{1}{2}\dot{q}^2(t)-\frac{1}{2}\omega^2q^2(t)+q(t)F(t).
\end{equation}
By considering (29) in conjunction with an associated equation
\begin{equation}
 \ddot{y}(t)+\omega^2y(t)=F(t)
\end{equation}
we can write a Lagrangian
\begin{equation}
 L_i=\dot{q}(t)\dot{y}(t)-\omega^2q(t)y(t)+q(t)F(t)+y(t)F(t)
\end{equation}
for the indirect analytic representation of the system. The simple method presented above is quite general and can also be used to deal with other non-homogeneous equations which appear in refs. 22, 23, 24 and similar studies.
\par {Lagrangians are called standard if they can be expressed as differences between \textquoteleft kinetic energy terms\textquoteright and \textquoteleft potential energy terms\textquoteright. The standard Lagrangians, in general, do not depend explicitly on time. In some cases, however, the associated Lagrangians may depend explicitly on time through exponential factors. The damped harmonic oscillator provides a typical example in this respect. The other forms of Lagrangians are referred to as non-standard ones. Linear Newtonian systems require to satisfy Helmholtz conditions to have a standard Lagrangian representation. The violation of Helmholtz conditions, however, does not provide a constraint for having non-standard Lagrangian representations of such systems \cite{25}. The work of Chandrasekar et al. \cite{26} provides a typical example in respect of this although their results can be obtained by using a relatively simple mathematical approach \cite{27} and although  similar works appear to have an old root in the  classical-mechanics literature \cite{28,29}. However, it is remarkable that the work in ref. 26 was envisaged  before the term \textquoteleft non-standard Lagrangian\textquoteright was coined by Musielak \cite{30} and some of the results of Chandrasekar et al. for Hamiltonian functions  were re-derived by  Bender et al. \cite{31} without taking recourse to the use of the extended Prelle-Singer method \cite{32,33}}.
\par We recognize that studies in Lagrangian and Hamiltonian structures of mechanical systems starting from one constant of the motion or first integral of the corresponding equation of motion are of considerable current interest \cite{14,26,34}. In this context, an interesting question arises: Can the equations of motion themselves, rather than their first integrals, be used to provide Lagrangian description of mechanical systems? This important issue was considered by Hojman \cite{35} who derived a very satisfactory method to obtain Lagrangian representation of linear Newtonian systems using their equations of motion. The basic philosophy of his approach is based on the following symmetry consideration.
\par {The relation between symmetries of a Lagrangian and conserved quantities of the corresponding equation of motion is provided by the so-called Noether's theorem  and  is a very well known result in classical mechanics. In contrast to this, it is less well known that the symmetries of the equations of motion form a larger set than the symmetries of the Lagrangian. However if  s-equivalence (a Lagrangian symmetry in which several constants of the motion  may be associated with one symmetry transformation) is taken into account the set of Lagrangian symmetries coincides with that of the equation of motion \cite{36}. This is perhaps the reason  why Hojman \cite{35} could find a short circuit to construct Lagrangians from the equations of motion rather than taking recourse to the use of their first integrals. However, instead of going into the details of the work in ref. 35, we shall apply the method to  construct a Lagrangian representation for a system of explicitly velocity-dependent two-dimensional differential equations given by
\begin{equation}
\ddot{x}+\gamma\dot{x}+\omega^2x=\alpha y
\end{equation}
and
\begin{equation}
\ddot{y}-\gamma\dot{y}+\omega^2y=\alpha x.
\end{equation}
Here $x=x(t)$ and $y=y(t)$. Henceforth we shall follow this notation throughout. Physically, (33) and (34) represent two coupled Harmonic oscillators embedded in a dissipative medium of frictional coefficient $\gamma$. Here $\alpha$ stands for the coupling constant. In the language of Bateman \cite{21} (33) represents a damped oscillator while the oscillator in (34) is an amplifier that absorbs energy drained out from that in (33). For the coupled system (33) and (34), we introduce the Lagrangian
\begin{equation}
L=y(\ddot{x}+\gamma\dot{x}+\omega^2x-\alpha y)+x(\ddot{y}-\gamma\dot{y}+\omega^2y-\alpha x)-\frac{d}{dt}(y\dot{x}+x\dot{y})
\end{equation}
which finally gives
\begin{equation}
L=\dot{x}\dot{y}+\frac{1}{2}\gamma(x\dot{y}-y\dot{x})+\frac{1}{2}\alpha(x^2+y^2)-\omega^2xy.
\end{equation}}
It is straightforward to see that (36) provides an indirect analytic representation of our coupled system and also that the system cannot have a direct analytic representation. More significantly, one can verify that if both equations  in (33) and (34) would represent damped oscillators, it would not be possible to construct an analytic representation for the coupled equations. Curiously enough, the undamped system corresponding to (33) and (34) can have both direct and indirect analytic representations \cite{11}.
\par So far we have studied the direct and indirect analytic representations of linear Newtonian systems in some detail.  We shall now  envisage a similar study for nonlinear systems with a view to illustrate the points of contrast and of similarity between the linear and nonlinear problems in respect of their analytic representations.  In close analogy with the works in refs. 11, 12 and 14 we begin the next section with a two dimensional autonomous differential equation (linear or nonlinear) and provide a general method to  find its Lagrangian from the constant of the motion. We then apply this approach to a number of physically important nonlinear equations.
\section*{5. Lagrangians for second-order nonlinear differential equations}
The general form of a two-dimensional differential equation (linear or nonlinear) can be written as
\begin{equation}
 \ddot{x}_i=f_i(x_j,\dot{x}_j),\;\;\;\;\;i,j=1,2.
\end{equation}
Equivalently, (37) reads
\begin{equation}
 \frac{dv_i}{dt}=f_i(\vec{x},\vec{v})
\end{equation}
with
\begin{equation}
 {\vec{x}}=(x_1,x_2),\;\;\;\;\; {\vec{v}}=\frac{d{\vec{x}}}{dt}=(v_1,v_2).
\end{equation}
A constant of the motion, $K(\vec{x},\vec{v})$,  of (37) or (38) satisfies 
\begin{equation}
 \sum_{i=1}^2(f_i(\vec{x},\vec{v}))\frac{\partial K}{\partial v_i}+v_i\frac{\partial K}{\partial x_i}=0
\end{equation}
along the integral curve of the equation. The solutions or integral surfaces of (40) can be obtained from the equation of characteristics \cite{15}
\begin{equation}
 \frac{dv_1}{f_1(\vec{x},\vec{v})}=\frac{dv_2}{f_2(\vec{x},\vec{v})}=\frac{dx_1}{v_1}=\frac{dx_2}{v_2}
\end{equation}
and subsequently used to write the Lagrangian of (37) as an integral representation \cite{11,12,14}
\begin{equation}
 L(\vec{x},\vec{v})=\frac{1}{2}\sum_{i=1}^2v^i\int^{v_i}\frac{K_n^{(i)}(\vec{x},\xi)}{\xi^2}d\xi.
\end{equation}
The subscript $n$ in $K$ has been used to differentiate between various constants of the motion that result from (41). Understandably, (42) represents an integral representation of the Lagrangian function. 
\par Let us now make use of (41) and (42)  to construct an indirect analytic representation for the  nonlinear equation
\begin{equation}                                                                                                                                                                                                                                                   
\frac{d^2x}{dt^2}+\sin x=0.                                                                                                                                                                                                                                                   \end{equation}
To that end we consider (43) together with an associated equation
\begin{equation}
\frac{d^2y}{dt^2}+\sin y=0. 
\end{equation}
Admittedly, this amounts to doubling the number of degrees of the system represented by (43). For (43) and (44) we can write the equation of characteristics as
\begin{equation}
 \frac{dv_x}{-\sin x}=\frac{dv_y}{-\sin y}=\frac{dx}{v_x}=\frac{dy}{v_y}.
\end{equation}
Equation (45) can be arranged in two different ways to get
\begin{equation}
 v_xdv_x+v_ydv_y=-\sin xdx-\sin ydy
\end{equation}
and
\begin{equation}
 v_xdv_y+v_ydv_x=-\sin xdy-\sin ydx.
\end{equation}
We can integrate (46) to obtain the constant of the motion
\begin{equation}
 K_1(\vec{x},\vec{v})=\frac{1}{2}(v_x^2+v_y^2)-\cos x-\cos y.
\end{equation}
Following the prescription given in ref. 11 we can construct from (48) the results for $K_1^{(1)}(.)$ and $K_1^{(2)}(.)$ which occur in (42). We thus obtain
\begin{equation}
 L=\frac{1}{2}(v_x^2+v_y^2)+\cos x+\cos y
\end{equation}
{to provide a direct analytic representation for nonlinear oscillator system (43) and (44). One would, therefore, expect that  the constant of the motion found from (47) will lead to an indirect analytic representation. But unfortunately, this is not possible because (47) cannot be integrated to write an expression for the constant of the motion. The simple example considered here concludes that the indirect analytic representation of an uncoupled nonlinear system cannot be found by doubling the number of degrees of freedom. One can verify that this conclusion is true for arbitrary nonlinear equations.
In the small angle limit $(\sin x=x,\;\sin y=y)$ our nonlinear system becomes identical to the linear system as given in (1) and (2). In this case (47) can be integrated to find a constant of the motion}
\begin{equation}
 K_2(\vec{x},\vec{v})=v_xv_y+xy
\end{equation}
to obtain an indirect analytic representation as noted in (4).
\par {In the above context we note that Chandrasekar et al. \cite{37} while  predicting some unusual nonlinear properties of the Li\'{e}nard-type oscillator 
\begin{equation}
 \ddot{x}+kx\dot{x}+\frac{k^2}{9}x^3+\lambda_1x=0
\end{equation}
made use of the extended Prelle-Singer method to obtain a direct Lagrangian for it. We demonstrate that a relatively simpler analytic representation of (51) can be found by writing it in the autonomous form
\begin{equation}
v(x)\frac{d}{dx}v(x)+kxv(x)+\frac{k^2}{9}x^3+\lambda_1x=0,\;\;\;\;\;v(x)=\frac{dx}{dt}.
\end{equation}
The result for the first integral of (51) can be obtained by solving (52) to read
\begin{equation}
 K_1(x,v)=h/g
\end{equation}
with
\begin{equation}
h=(9\lambda_1+k^2x^2+3kxv)^2
\end{equation}
and
\begin{equation}
 g=9\lambda_1+k^2x^2+9kxv.
\end{equation}
From (42) and (53) we get the required Lagrangian as
\begin{equation}
L=3k\dot{x}\log (9\lambda_1+k^2x^2+6k\dot{x})-2k^2x^2.
\end{equation}}
The result in (56) provides a direct analytic representation for Li\'{e}nard-tytpe oscillator. Since (51) involves a velocity-dependent or dissipative term, it may be of some interest to follow  Bateman \cite{21} and introduce a dual system
\begin{equation}
\ddot{y}-ky\dot{y}+\frac{1}{9}k^2y^3+\lambda_1y=0
\end{equation}
with a view to look for an indirect analytic representation for the equation in (51). We have verified that, as opposed to the linear dissipative systems, no such representation exists for the Li\'{e}nard  type oscillator and this is true for other uncoupled nonlinear dissipative equations.
\par We next consider the coupled Duffing oscillators given by \cite{38} 
\begin{equation}
\ddot{x}+\omega^2x+4\alpha x^3+12\alpha xy^2=0\;\;\;\;\mbox{and}\;\;\;\;\ddot{y}+\omega^2y+4\alpha y^3+12\alpha x^2y=0
\end{equation}
{which represent an important nonlinear system that plays a role in many applicative contexts including the detection of machinery faults \cite{39}. For the differential equations in (58), the equation of characteristics (41) can be arranged in two different ways so as to obtain the following results for the constants of the motion given by
\begin{equation}
K_1=\frac{1}{2}(1+c_1^4)\dot{x}^2+\frac{\omega^2}{2}(x^2+y^2)+\alpha(x^4+y^4)+6\alpha x^2y^2
\end{equation}
and
\begin{equation}
K_2=c_1^2\dot{x}+\omega^2xy+4\alpha x^3y+4\alpha xy^3,
\end{equation}
where $c_1^2=\dot{y}/\dot{x}$ that remains constant while constructing the corresponding Lagrangians by the use of (59) and (60) in (42). The equations (59) and (42) give the direct Lagrangian
\begin{equation}
L_d=\frac{1}{2}(\dot{x}^2+\dot{y}^2)-\frac{\omega^2}{2}(x^2+y^2)-\alpha(x^4+y^4)-6\alpha x^2y^2
\end{equation}
for the coupled system in (58). On the other hand, (60) and (42) lead to
\begin{equation}
L_i=\dot{x}\dot{y}-\omega^2xy-4\alpha x^3y-4\alpha xy^3,
\end{equation}
the so-called indirect Lagrangian of the system.
\par  In close analogy with (33) and (34) for the coupled damped Harmonic oscillators we introduce the damped Duffing oscillators as
\begin{equation}
\ddot{x}+\gamma\dot{x}+\omega^2x+4\alpha x^3+12\alpha xy^2=0\;\;\;\;\mbox{and}\;\;\;\;\ddot{y}-\gamma\dot{y}+\omega^2y+4\alpha y^3+12\alpha x^2y=0
\end{equation}
and verify that the system in (63) can  have  the indirect analytic representation only given by
\begin{equation}
L_i=\dot{x}\dot{y}-\omega^2xy-\frac{\gamma}{2}(\dot{x}y-x\dot{y})-4\alpha(x^3+xy^3).
\end{equation}
The system of equations in (63) has been used to model Soret driven Benard convection \cite{40}, vibration of stretched string \cite{41} and motion of nonlinear circular plates \cite{42}}.
\section*{6. Concluding remarks}
Representation of dynamical systems by Lagrangians or the so-called analytic representation plays a key role in  diverse areas of physics ranging from classical mechanics to  quantum field theory. In general, for any given system one can solve the inverse variational problem to construct either the direct or indirect analytic representation. Mechanical systems can also  admit both representations simultaneously. The most common example in respect of this is provided by the damped harmonic oscillator. In this work we have explicitly demonstrated that self-adjoint form of the equation of motion provides a basis to construct direct analytic representation. On the other hand, the original Newtonian  equation  and  its adjoint taken together can be used to derive the so-called indirect analytic representation. Our approach to the inverse problem clearly shows that how, without taking recourse to the use of phenomenological arguments, one can efficiently employ  the theory of second-order differential equation to bring open systems within the framework of the action principle. We have first examined this by dealing with the  damped harmonic oscillator and then presented a number of case studies.
\par  In recent years there has been resurgence of interest in the Lagrangian and Hamiltonian  description of dissipative systems \cite{43,44}. The canonical quantization of damped Harmonic oscillator using the indirect Lagrangian representation have been found to be quite straightforward \cite{45,46,47} because the corresponding Hamiltonian is time independent. The Hamiltonians corresponding to the results for indirect Lagrangians in Table II are, however, not time independent. Thus it will be interesting to derive a quantization procedure for systems characterized by time-dependent indirect Lagrangians.
\par We noted that it is not straightforward to use the theory of nonlinear self-adjointnes \cite{9,10} for solving the inverse variational problem of nonlinear differential equations. We thus made use of an integral representation of the Lagrangian function \cite{14} to compute results for the direct and indirect Lagrangians of a number of physically important nonlinear systems. Interestingly, we found that, as opposed to the result obtained for an uncoupled linear equation, the indirect Lagrangian for the corresponding nonlinear equation cannot be constructed by doubling the number of degrees of freedom of the system. On the other hand, the coupled equations, whether linear or nonlinear, can have both direct and indirect analytic representations. The corresponding dissipative systems, however, follow from indirect Lagrangians only. If the time evolution of a mechanical system is governed by linear differential equations, ordinary or partial, the solution of the problem can be studied confidently because linear analysis is based on the assumption that individual effects can be unambiguously traced back to particular causes. This assumption does not hold good for nonlinear analysis such that patching simple pieces to understand the whole simply does not work. In fact, as compared to linear systems, the causal information flow in nonlinear ones is highly complicated \cite{48}. It may, therefore, be tempting to attribute the observed anomaly between the analytic representations of linear and nonlinear equations (uncoupled) to the difference in causal relation in these systems. But it remains an interesting curiosity to provide an explanation for why the analytic representations of both linear and nonlinear coupled system exhibit identical behavior. Moreover, one may also like to extend our method to construct Lagrangians for isochronous Li\'{e}nard-type oscillators of different dimensions \cite{49,50}.


\vskip 1cm
\begin{thebibliography}{99}
\bibitem{1} R M Santilli, {\it Foundation of Theoretical Mechanics}, Vol. 1- {\it The inverse Problem in Newtonian Mechanics}, Springer Verlag, New York, 1978)
\bibitem{2} J Douglas, Trans. Amer. Math. Soc. {\bf 50}, 71 (1941) 
\bibitem{3} M Crampin, T Mestdag and W Sarlet, ZAMM Z. Angew. Math. Mech. {\bf 90}, 502 (2010)
\bibitem{4} H Goldstein, {\it Classical Mechanics}, (Addison-Wesley, Reading, MA, 1950).
\bibitem{5} G Morandi, C Ferrario, G Lo Vecchio, G Marmo and C  Rubano, Phys. Rep. {\bf 188}, 147 (1990)
\bibitem{6} R Courant and D Hilbert, {\it Methods of Mathematical Physics}, Vol. 1 (Wiley Eastern Pvt. Ltd., New Delhi, 1975).
\bibitem{7} E A Coddington and N Levinson: {\it Theory of Ordinary Differential Equations} (McGraw-Hill Book Co., New York, 1955)
\bibitem{8} E L Ince, {\it Ordinary Differential Equations} (Dover Publications, New York, 1958). 
\bibitem{9} N H Ibragimov, J. Math. Anal. Appl. {\bf 318}, 742 (2006)
\bibitem{10} N H Ibragimov, J. Phys. A: Math. Theor. {\bf 44}, 432002 (2011)
\bibitem{11} S Ghosh, B Talukdar, P Sarkar and U Das, Acta Mechanica {\bf 190}, 73 (2007)
\bibitem{12} A Saha and B Talukdar, Rep. Math. Phys. {\bf 73}, 299 (2014)
\bibitem{13} Rami Ahmed El-Nebulsi, Can. J. Phys. {\bf 93}, 55 (2015)
\bibitem{14} G L\'{o}pez, Ann. Phys. (NY) {\bf 251}, 363 (1996)
\bibitem{15} I N Sneddon, {\it Elements of Partial Differential Equations} (McGraw-Hill Book Co., NY, 1957). 
\bibitem{16} C Caratheodory, {\it Calculus of variations and partial differential equations of the first order}, Vol. 2 (Holden-Day  Inc., San Fransisco, 1967)
\bibitem{17} B Talukdar and U Das, {\it Higher-Order Systems in Classical Mechanics}, ( Narosa Publishing House, New Delhi, 2008)
\bibitem{18} P Caldirola, Il Nuovo Cimento {\bf 18}, 393 (1941)
\bibitem{19} E Kanai, Prog. Theor. Phys. {\bf 3}, 440 (1948)
\bibitem{20} Z E Musielak, N Davachi and M Rosario-Franco, Mathematics {\bf 8}, 379 (2020)
\bibitem{21} H Bateman, Phys. Rev. {\bf 38}, 815 (1931) 
\bibitem{22} M G Fuda and J S Whiting, Phys. Rev. C {\bf 8}, 1255 (1973) 
\bibitem{23} B Talukdar, M N Sinha Roy, N Mallick and D K Nayek, Phys. Rev. C {\bf 12}, 370 (1975)
\bibitem{24} B Talukdar, N Mallick and S Mukhopadhyaya, Phys. Rev. C {\bf 15}, 1252 (1977)
\bibitem{25} Z E Musielak, Chaos, Solitons and Fractals {\bf 42}, 2645 (2009)
\bibitem{26} V K Chandrasekar, M Senthivelan and M Lakshmanan, J. Math. Phys. {\bf 48}, 032701 (2007)
\bibitem{27} Subrata Ghosh, J  Shamanna and B Talukdar, Can. J. Phys. {\bf 82}, 561 (2004)
\bibitem{28} R de Ritis, G Marmo, G Plastino and P Scudellaro, Int. J. Theor. Phys. {\bf 22}, 931 (1983)
\bibitem{29} P Havas, Il Nuovo Cimento {\bf 5}, 363 (1957)
\bibitem{30} Z E Musielak, J. Phys. A: Math. Theor. {\bf 41}, 055205 (2008)
\bibitem{31} C M Bender, M Gianfreda, N Hassanpour and H F Jones, J. Math. Phys. {\bf 57}, 084101 (2016)
\bibitem{32} M J Prelle and M F Singer, Trans. Am. Math. Soc. {\bf 279}, 215 (1983)
\bibitem{33} V K Chandrasekar, M Senthilvelan and M Lakshmanan, Proc. Roy. Soc. A {\bf 461}, 2451 (2005)
\bibitem{34} S A Hojman, Acta Mech. {\bf 226}, 735 (2015)
\bibitem{35} S A Hojman, J. Phys. A: Math. Gen. {\bf 17}, 2399 (1984)
\bibitem{36} R Hojman, S A Hojman and J Sheinbaum, Phys. Rev. D {\bf 28}, 1333 (1983)
\bibitem{37} V K Chasndrasekar, M Senthilvelan and M Lakshmanan, Phys. Rev. E {\bf 72}, 066203 (2005)
\bibitem{38} L D Landau and E M Lifshitz, {\it Mechanics} (Pergamon Press, New York, 1982)
\bibitem{39} N Q  Hu and X S Wen, J. Sound and Vibration {\bf 268}, 917 (2003)
\bibitem{40} M A F Sanju\'{a}n, J. L. Valero and M G Velarde, Il Nuovo Cimento D {\bf 13}, 913 (1991)
\bibitem{41} J A Elliott, Am. J. Phys. {\bf 50}, 1148 (1982)
\bibitem{42} T A Nayfeh and A F  Vakakis, Int. J. Non-Linear Mech. {\bf 29}, 233 (1994)
\bibitem{43} J. Guerrero, F F L\'{o}pez-Ruiz, V Aldaya and F Cass\'{i}o, J. Phys. A : Math, Theor. {\bf 44}, 445307 (2011)
\bibitem{44} N E Mart\'{i}nez-Perez and C Ram\'{i}rez, J. Math. Phys. {\bf 59}, 032904 (2018)
\bibitem{45} M Blasone and P Jizba, Ann. Phys. (NY) {\bf 312}, 354 (2004)
\bibitem{46} K Takahashi, J. Math. Phys. {\bf 59}, 032103 (2018)
\bibitem{47} K Takahashi, J. Math. Phys. {\bf 59}, 072108 (2018)
\bibitem{48} S Li, Y Xiao, D Zhou and D Cai, Phys. Rev. E {\bf 97}, 052216 (2018)
\bibitem{49} M Sabatini, J. Diff. Eqns. {\bf 152}, 467 (1999)
\bibitem{50} A K Tiwari, A Durga Devi, R Gladwin Pradeep and V K Chandrasekar, Pramana, J. Phys. {\bf 85}, 789 (2015)
\end{thebibliography}
\end{document}